# Multistate polarization addressing using one single beam in an azo polymer film

S. Ahmadi Kandjani [1,2], R. Barille [1], S. Dabos-Seignon [1], and J.-M. Nunzi [1],

[1] *Laboratoire POMA*
*UMR 6136 - Université d'Angers*
*4, Boulevard Lavoisier, B.P 2018,*
*49016 Angers, France*

[2] *Research Institute For Applied Physics and Astronomy ( RIAPA ),*
*The University of Tabriz, 51664 Tabriz, Iran*

E. Ortyl, S. Kucharski
*Institute of organic and polymer technology*
*Wroclaw Technical University,*
*50-370, Wroclaw, Poland*

**Abstract**

Peculiar light-matter interactions can break the rule that a single beam polarization can address only two states in an optical memory device. Multistate storage of a single beam polarization is achieved using self-induced surface diffraction gratings in a photo-active polymer material. The grating orientation follows the incident light beam polarization direction. The permanent self-induced surface relief grating can be readout in real time using the same laser beam.

# Multistate polarization addressing using one single beam in an azo polymer film

S. Ahmadi Kandjani, R. Barille, S. Dabos-Seignon, J.-M. Nunzi, E. Ortyl, S. Kucharski

Owing to the fact that photon possesses a polarization angular momentum of one, it is well admitted that one single beam polarization can address only two states in an optical memory device. Indeed, use of a polarized light can in principle only double the storage capacity of digital or holographic optical memory devices [1]. We show in the following that peculiar material properties can break this principle.

Properties of polymers containing photochromic azo dyes have received tremendous attention owing to the possibility of optically addressed birefringence and dichroism [2, 3]. This led to the investigation of the holographic optical storage properties of azo polymers [4]. It is now well-established that following light excitation whose wavelength lies in the absorption band, cis-trans isomerization takes place, leading to further thermal orientation diffusion which enables a molecular rotation within the polymer matrix. This finally leads to a full reorientation of the azo-dye molecules [5, 6]. More recent results showed that surface relief gratings (SRG) were also induced further to the photo-excitation of azo-polymer films [4, 7]. Photoinduced SRG were interpreted as a consequence of molecular translation controlled by the optical field [8]. An interference pattern of coherent light beams was used for irradiation of the materials and the film surface modification was controlled by the light interference pattern [9]. Two beams are usually necessary but it was also recently demonstrated that a single beam interaction can induce well ordered structures [10].

In the same time, the increase of information storage needs in information science and technology lead to significant efforts in order to increase the capacity of storage media. Holographic memories are promising candidates in this respect because they permit 3-dimensional optical information storage [11]. In this case, coding is done by the mixing of two coherent laser beams: an object and a reference one. The angular selective property of holograms recorded in thick materials enables high density data storage. Two beams are usually necessary. Photorefractive materials are attractive for hologram recording, but they are weakly sensitive to the intensity distribution of the recording beam and require 2 beams for a multidimensional addressing [12].

In this study we demonstrate that multistate storage of a single beam polarization can be achieved and readout in a polymer material. Our original technique uses only one beam with controlled polarization in order to photoinduce a SRG whose wave-vector direction depends on the light polarization. The trial material used in this study is an efficient azo-polymer derivative [13].

SRG are produced in a one step irradiation process in standard laboratory conditions. Our characterization method for photoinduced SRG consists in diffraction studies and atomic force microscopy (AFM). Our observations open the door to multistate optical information storage using polarization encoding.

The samples are polymer films made from a highly photoactive azobenzene derivative containing heterocyclic sulfonamide moieties: 3-[{4-[(E)-(4-{[(2,6-dimethylpyrimidin-4-yl) amino] sulfonyl}phenyl) diazenyl]phenyl}-(methyl)amino]propyl 2-methylacrylate [14]. Thin films on glass substrates were prepared by spin-coating of the polymer from THF solutions with a concentration of 50 mg/ml. Thickness measured with a Dektak-6M Stylus Profiler was around 1μm. Absorbance at λ = 438 nm maximum is 1.9 The λ = 476.5 nm laser line of a continuous argon ion laser is used to excite the azo polymer absorption close to its absorption maximum. Absorbance at working wavelength is 1.6. Incoming light intensity is controlled by the power supply. Polarization direction of the laser beam is varied using a half-wave plate. Sample is set perpendicular to the incident laser beam. The size of the collimated laser beam impinging onto the polymer sample is controlled with a Kepler-type afocal system. Sample is irradiated with different polarizations using different laser beam intensities with a defect beam size of 4 mm diameter at $1/e^2$.

In a preliminary experiment we carefully checked that laser irradiation leads to a topographic modification of the polymer surface resulting in a SRG. When the surface relief is formed, the impinging beam itself is diffracted in several diffraction orders. We show in figure 1 the intensity for first order self-diffraction recorded as a function of time for different laser beam intensities. Self-diffraction which occurs in both forward and backward directions is collected in the backward direction by a f = 200 mm focal-length lens and registered as a function of time by a photodiode. Above 760 mW/cm$^2$ input beam intensity, the polymer film could be damaged and we did not exceed this limit. The self-diffraction phenomenon exhibits a threshold depending on the power density. We define a threshold time for the induction of self-diffraction as the inflexion point of the curve, when second derivative changes sign. The threshold is linear as a function of the input beam intensity and its slope is estimated at about 33 mW/cm$^2$.min$^{-1}$. We checked that this threshold does not depend on polarization. Measurements using different beam sizes between 3 and 6 mm confirmed also that the threshold time is a function of power density only. For a same laser intensity but with a smaller beam size, threshold time decreases.

We then controlled at a microscopic level the grating formation as a function of time (fig. 2). In order to get information on the surface relief evolution during the recording period, we have printed different gratings on the same sample during different times with a laser intensity of 450 mW/cm$^2$.

Both the height and the pitch of the grating were retrieved as a function of time with a contact-mode AFM. Insert in figure 2 presents the first order self-diffraction intensity as a function of time up to the saturation intensity at which a stable and permanent grating is printed. All results in figure 2, for a 450 mW/cm$^2$ intensity, either for the grating's height or pitch and diffraction show a threshold-like evolution with inflexion around 20 min.

Figure 3 (a-d) shows AFM images of permanent structures induced with four different polarizations of the input laser beam: at 0°, 30°, 60° and 90° with respect to the initial polarization. Laser beam intensity was 450 mW/cm$^2$. Growth time for these gratings was 1h. It corresponds to the first order diffracted beam intensity reaching a maximum. AFM measurements retrieve angle values which are in a good agreement with the input polarization direction. Surface gratings have a depth of 50 nm ± 5 nm, whatever the polarization used. Grating pitch is Λ = 800 nm ± 30 nm, which is in agreement with the value given by second order diffraction theory in the θ = 32.6° direction: Λ = 2 λ / 2 sin θ. Figure 3 (e-h) shows camera pictures of the self-diffracted beam for each polarization angle. Measurements using different beam sizes between 3 and 6 mm confirmed that the time to reach a permanent grating is a function of power density only.

The mechanism proposed to explain the origin of the driving force responsible for SRG formation was partly disclosed earlier [15]. It results from isomerization induced translation in which the molecules migrate almost parallel to the polarization direction [8]. In particular, when a collimated laser beam with intensity $I$ impinges onto the sample, light is diffused inside the polymer film in all directions around any micro-roughness [15]. Let's call $h$ the height of a particular roughness. It will diffract a light amplitude $u$ into the film plane with $u \propto hA$ where $A = I^{1/2}$ is the laser amplitude. According to [8], the rate of growth of the roughness is $\partial h/\partial t \propto u^+ \cdot u^-$ where $u^+$ and $u^-$ stand for two counter propagating coherent waves interfering into the polymer film plane. So, roughness height $h$ increases with time $t$ from its initial value $h_0$ as $h(t) \approx h_0 / (1 - b I t h_0)$, up to saturation of the diffraction efficiency owing to coupled wave theory [16]. Coefficient $b$ is related to the quantum efficiency of the photoinduced translation and surface relief growth process [8]. The process which initially diverges as $b I t h_0$ tends to 1 saturates and tends to a limit height $h_{max}$ for which more light coupled by diffraction into the polymer film results into more light diffracted out of the polymer film. A balance is then obtained when incident and diffracted beams reach comparable magnitudes (backward up-diffracted intensity reaches experimentally 1 %).

When the grating has been stored permanently, the self-induced SRG can be retrieved using the same beam as the one used to write the grating. Before saturation of the SRG ($t < 20$ min in fig.2), the process is reversible and diffraction can be addressed in several direction using the writing beam polarization. This shows on one side that the fundamental process of SRG inscription in our azo-polymer is reversible. This shows on an other side that several diffraction directions can be controlled using one single beam polarization direction. The minimum number of diffraction directions which can be resolved without overlapping is limited by the lateral size of the first order spots in the Fourier transform of the SRG in figure 3-left. This Fourier transform is exactly given by the diffraction pattern shown in figure 3-right. This practically results in a divergence of the diffracted beam: $\alpha = 22.6°$. If we detect the diffracted beam at a distance $D$ from the polymer film, the spot diameter is: $d = 2D \tan \alpha/2$. So, the maximum number of states $N$ that can be encoded and readout without overlap is:

$$N = \frac{S}{2d} = \frac{\pi \sin \theta}{2 \tan \alpha/2}$$

where $S$ is the perimeter of the circle made by the diffraction angle in the detection plane. The factor 2 in the denominator accounts for negative and positive diffraction orders. We find $N = 4$ in our particular example.

We have demonstrated for the first time that multistate addressing by polarization could be achieved using one single beam polarization into an azo-polymer film. The different polarization states are stored as surface diffraction gratings. Grating orientation follows the incident beam polarization direction. The permanent SRG readout by a single beam diffracts light into positive and negative diffraction orders in the backward and forward directions along the incident beam polarization direction. We are currently developing faster materials in order to find realistic applications of our principles in optical processing devices.

**FIGURE CAPTIONS**

Figure 1: Intensity of the first order of diffraction measured as a function of time for different laser beam intensities. The straight line for I = 560 mW/cm$^2$ estimates the time threshold value for the SRG recording.

Figure 2: Evolution of the height and pitch of the grating as a function of time at the beginning of the surface relief grating recording for a beam intensity of 450 mW/cm$^2$. The insert shows the first order diffraction intensity measured as a function of time up to the saturation of the recorded grating.

Figure 3: AFM images of typical surface grating displaying a SRG amplitude of 50 nm and pitch of 800 nm for four different input beam polarizations, (a) θ = 0°, (b) θ = 30°, (c) θ = 60°, (d) θ = 90°. Parts (e-h) show the first order self-diffracted images retrieved with the recording laser.

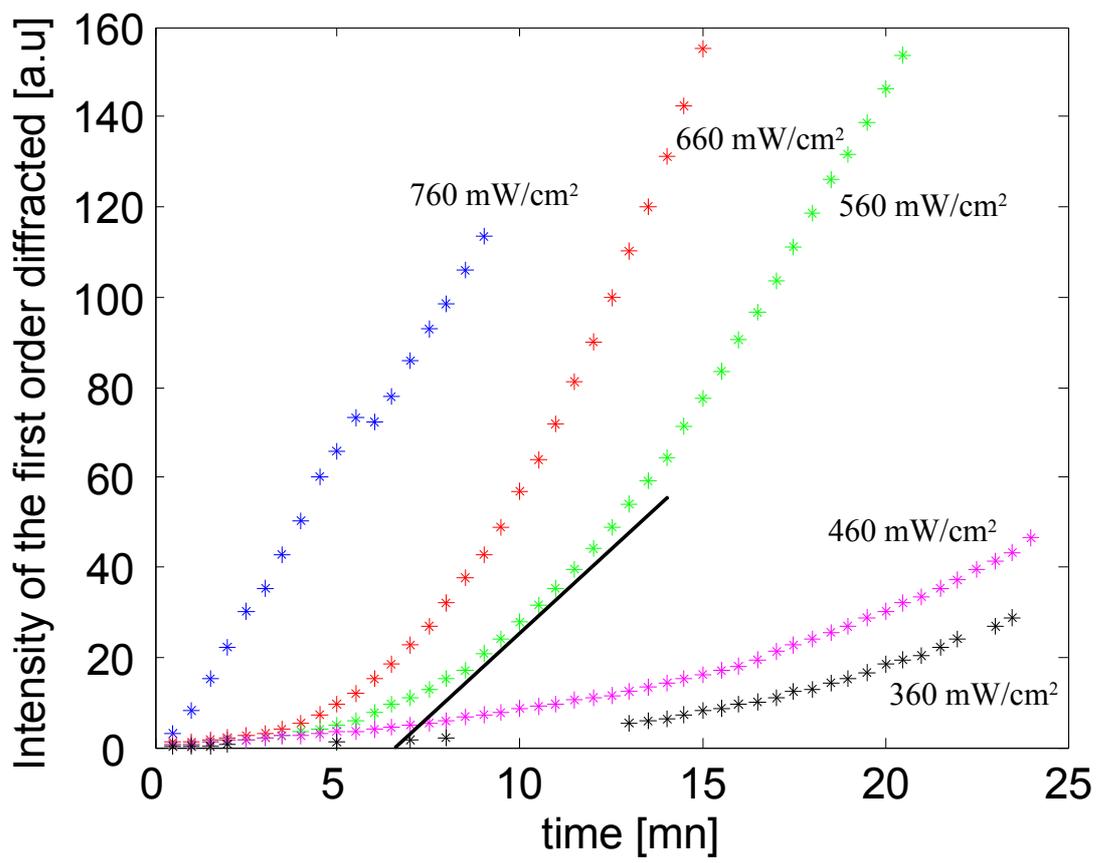

**Figure 1**

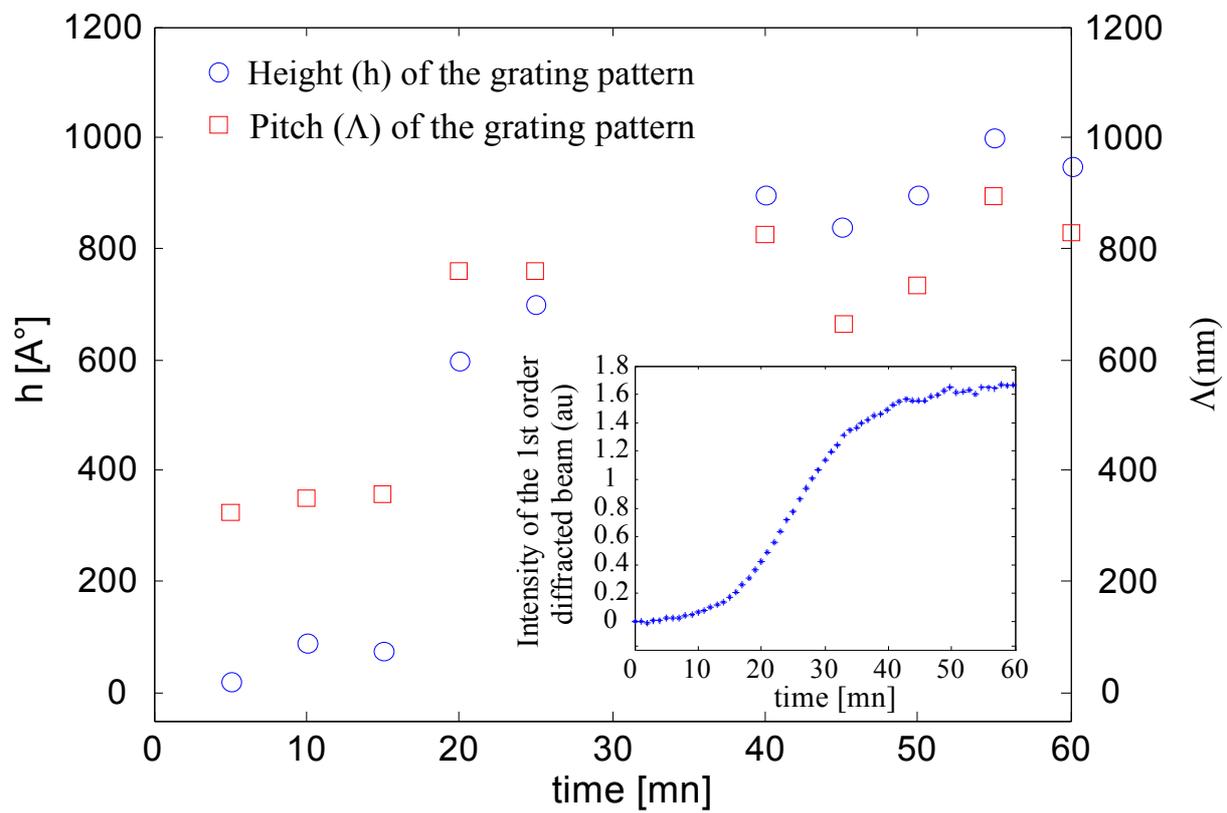

**Figure 2**

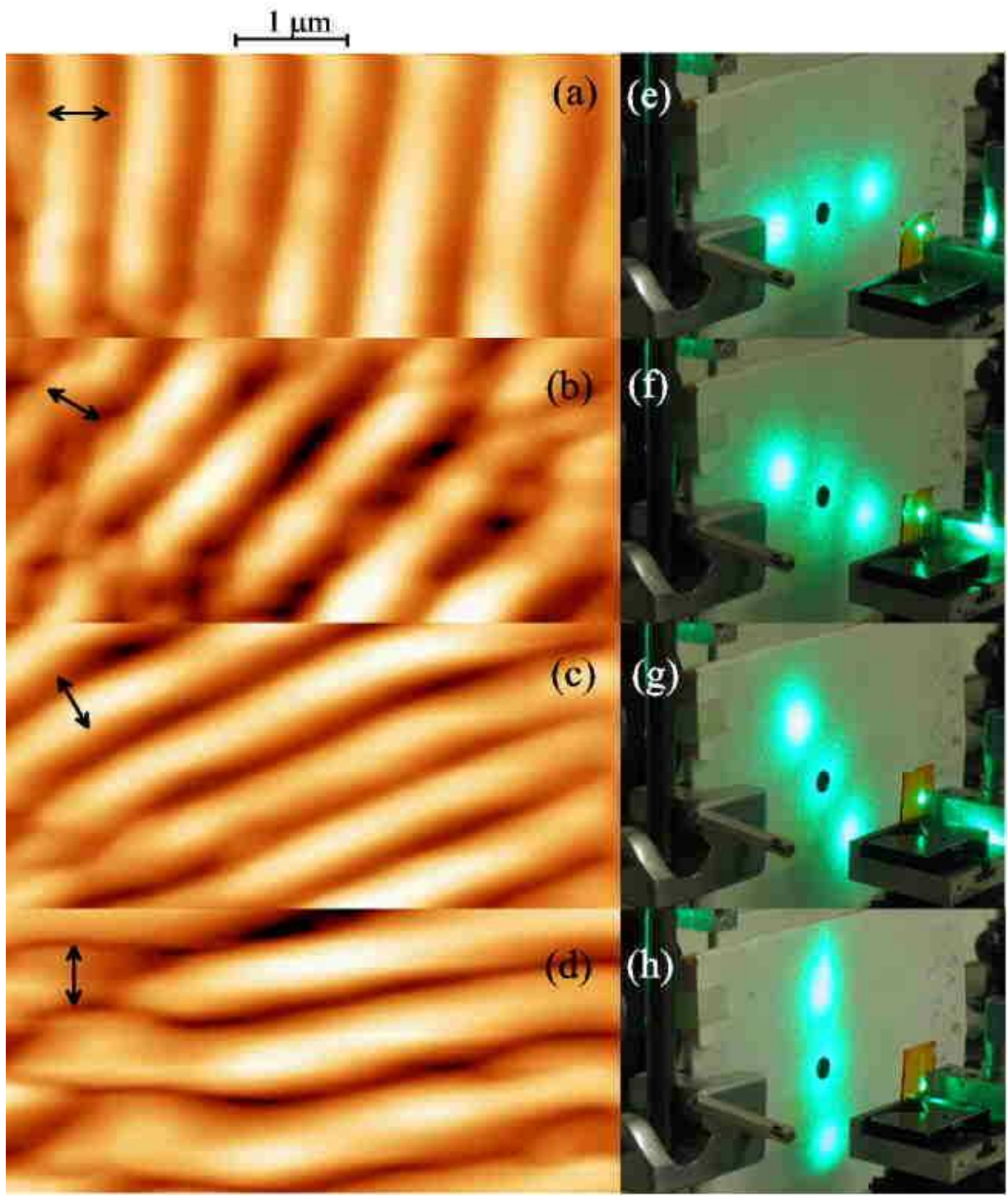

**Figure 3**